\documentclass[pre,aps,floats,superscriptaddress,floatfix,preprint]{revtex4}
\usepackage{amssymb,amsmath}
\usepackage{amsmath,amssymb}
\usepackage{graphicx}
\usepackage{psfrag}

\def\beq{\begin{equation}}
\def\eeq{\end{equation}}
\def\bea{\begin{eqnarray}}
\def\eea{\end{eqnarray}}
\begin{document}
\title{Phase transitions and continuously variable scaling in a chiral quenched disordered
model }
\author{Niladri Sarkar}\email{niladri.sarkar@saha.ac.in}
\author{Abhik Basu}\email{abhik.basu@saha.ac.in}
\affiliation{Theoretical Condensed Matter Physics Division, Saha
Institute of Nuclear Physics, Calcutta 700064, India}

\date{\today}

\begin{abstract}
We elucidate the effects of chiral quenched disorder on
the scaling properties of pure systems by considering a reduced
model that is a variant of the quenched disordered cubic anisotropic
$O(N)$ model near its second order phase transition. A generic
short-ranged Gaussian disorder distribution is considered. For
distributions not invariant under spatial inversion ( {hence
chiral}), the scaling exponents are found to depend continuously on
a model parameter that describes the extent of inversion symmetry
breaking. Experimental and phenomenological implications of our
results are discussed.
\end{abstract}


\maketitle

\section{Introduction}

 The large-scale, macroscopic effects of
disorder in statistical mechanics models and condensed matter
systems have been a subject of intense study for a long time now.
Very generally, depending upon time-scales, there can be two kinds
of structural disorders that can exist in a system, namely, annealed
and quenched disorders. In a system with annealed disorder the
impurities can diffuse freely in a system until they come to a thermal 
equilibrium state. The time-scale of the annealed disorder dynamics
is comparable or shorter than that for the physical degrees of
freedom of the corresponding pure system, and hence the dynamics of
the disorder becomes important. In contrast, for a system with a
quenched disorder the impurities are fixed in particular configurations 
and do not evolve in time, 
and, as a result, the disorder configuration is not in thermodynamic
equilibrium. Studies on the effects of random quenched disorder on
pure systems, e.g., the  {classical} $O(N)$  {spin} model
\cite{mw,bc,wh} and self-avoiding walks on random lattices
\cite{ck,harris,kim1,rc,kim2} are important, particularly because of
the modifications in the critical behaviour brought about by the
presence of such impurities.  Whether or not quenched disorder
changes the universal scaling properties of the pure system is given
by the heuristic arguments of the well-known {\em Harris criterion}
\cite{harris}. For Gaussian-distributed random impurities, both
spatially short-ranged \cite{lub} or long-ranged \cite{wh,kim2}
variances have been considered. For instance, perturbative
renormalization group (RG) calculations on $O(N)$ symmetric models
yield that for short-ranged disorder, the scaling exponents, even
when they are affected by the disorder, do not generally depend upon
the strength of the disorder~\cite{lub}. In addition, experiments on
liquid crystal systems reveal important (i.e., relevant in a RG
sense) effects of quenched disorder~\cite{exp1}. These studies
typically considered disordered media with variances invariant under
spatial parity inversion (achiral). In contrast, the influence of
chiral disordered media on the scaling properties of pure systems
are yet to be explored systematically. Chirality is known to affect
universal properties of systems; see, e.g., Refs.\cite{noise,noise1}
where effects of chirality is discussed on a class of driven
nonequilibrium models. Furthermore, it is now possible to fabricate
chiral liquid crystals by embedding nematic liquid crystals into
porous chiral film made of deposition of helical columns of MgF$_2$
on glass substrates. This is believed to be of use in practical
applications, e.g., optical switching~\cite{nema-chiral}. In view of
these examples and considering the fact that chiral disordered
substance are now prepared experimentally (e.g., chiral
aerogels~\cite{chiral-aero}), it is pertinent and relevant to
theoretically investigate the effects of chiral disordered media on
the scaling properties of pure systems in a general setting.

In this article, we propose a reduced minimal model with quenched 
chiral disorder and analyse 
it to study the generic effects of such a kind of disorder 
on the critical scaling of pure systems. While our
intention
 is not to model any specific real system
in details, our results should have broad implication in
understanding of media with orientational disorder, e.g., the effects of chiral porous media
\cite{chiral-aero} on the nematic-to-smectic A (N-A) transition of
liquid crystals or on the scaling properties of smectic-A liquid
crystals.  {Effects of usual (achiral) short-ranged disorder on N-A
transitions and smectic A are both theoretically and experimentally
well-studied~\cite{ben} and a variety of results are obtained
including possible destruction of translational (smectic) order, but
stabilisation of the smectic Bragg glass phase. There are other
systems where effects of chiral quenched disorder is likely to be of
significance, e.g., disordered cholesteric liquid
crystals~\cite{ben}, smectic A to C (A-C) transition~\cite{toner}
and superfluids in aerogels~\cite{volo}. The first one would be
particularly intriguing, due to the fact that  pure cholesteric
liquid crystals are themselves chiral. However, the ensuing
algebraic manipulations for the specific systems mentioned above {in
conjunction with chiral quenched disorder and its coupling with pure
system variables}, will be quite challenging. Thus, studies on a
simpler reduced minimal model should be welcome. Such approaches are
useful provided it allows one to address questions of basic
principles, which in the present case is the effects of chirality or
handedness of the disorder medium. To this effect,
 {in this article} we consider a variant of the usual
quenched disorder} $O(N)$  model ($N\geq 2$) with cubic anisotropy
 {in the presence of a disorder distribution that {\em
breaks the invariance under spatial reflection} or {\em parity
inversion} and study its scaling properties near the critical
point}. Our model is essentially a generalisation of the
random cubic anisotropic model introduced in Ref.~\cite{old}; see also Ref.~\cite{old1}
for related discussions on a problem of unconventional superconductors with
quenched impurities. The
lack of parity inversion symmetry in our model represents chirality or handedness
of the disordered medium. Possible macroscopic effects of the
chirality of the disordered medium, e.g., effects on universal
scaling properties have not been considered before; nevertheless,
{formally} their existence cannot be ruled out on any general
(microscopic) ground, for they represent lack of reflection
invariance of the impurity
distribution at small scales. 
By systematically using a perturbative RG framework together with
the replica formalism~\cite{grinlut,replica}, we find (a) for order
parameter component $N>4$, the disorder is {\em irrelevant} (in a RG
sense, see below) with the second order phase transition in the
model being described by the corresponding pure system scaling
exponents, (b) for $N<4$ when the disorder distribution is
reflection invariant, the system is described by a random isotropic
(RI) fixed point (FP), at which disorder is relevant (in a RG sense)
and which is identical to the scaling behaviour of the isotropic
$O(N)$ symmetric model with short-ranged quenched
disorder~\cite{lub}, in agreement with Ref.~\cite{old}, and
(c) one may vary this random isotropic FP {\em continuously} by
tuning a non-negative model parameter $N_\times$ (see below), that
describes the strength of the parity-breaking parts in the variance
of the disorder distribution from zero. The scaling exponents that
characterise the second-order phase transition at this FP vary
continuously with $N_\times$. At the technical level, as we shall
see below, $N_\times$ appears as a marginal operator in the ensuing
disorder-averaged state of the system. In addition, for a non-zero
$N_\times$, near the critical point the model displays a diverging
correlation length $\xi$ and a fluctuation-corrected {\em
renormalized} $T_c$ that depend on $N_\times$, suggesting the
possibility of $N_\times$-dependent scaling near renormalized
critical point in different realizations of the model system. Our
results should be directly testable in disordered classical systems
with carefully chosen distributions for the disorder at dimension
$d=3$~\cite{topo}. The remaining part of the article is organised as
follows: In Sec.\ref{model}, we propose and define our model. Then
we extract the critical scaling exponents from the model in
Sec.\ref{res}. In Sec.\ref{conclu}, we summarise and discuss our
results.

\section{The model}
\label{model}

Let us begin by discussing possible requirements of a simple model
that will be able to capture the effects of chiral quenched
disordered media. First of all, if the chirality of the disordered
medium has any discernible effect on the system, we expect two-point
correlation functions of the physical variables that describe the
corresponding pure system to display chirality (i.e., have parts
that are odd parity under spatial inversion). This evidently
requires two or more variables describing the disorder in the
system, since only a cross-correlation function of two different
variables can have a part that is odd under spatial inversion.
Secondly, for an impurity distribution that displays lack of
symmetry under inversion, it is expected to be described by more
than one (frozen) field. Keeping these features in mind, and
considering systems that are purely achiral in the absence of any
disorder, we consider a disordered version of the well-known pure
cubic anisotropic $O(N)$ model~\cite{lubensky-book} as a reduced
minimal model for chiral quenched disorder systems. The
corresponding free energy functional is $\mathcal {F}=\int d^dx f$
with
 \bea f =
\sum_{i=1}^N[{1 \over 2}\{r_0 (\phi_i^2) + r_i({\bf x})\phi_i^2 +
(\nabla\phi_i)^2 + u(\phi_i^2)^2 + v\phi_i^4], \label{free}
 \eea
where $r_0=(T-T_c)/T_c$ with $T$ and $T_c$ referring to the
temperature and the mean-field critical temperature, respectively.
Stochastic function $r_i({\bf x})$ represents the coupling with the
disorder, such that $T_c^L=T_c+r_i({\bf x})$ is the local {\em
fluctuating } critical temperature for $\phi_i$.  Further, $u>0$ and
$v>0$ are the bare nonlinear coupling constants in the model. In the
limit of $v=0$ and for all $ r_i({\bf x})= \tilde r({\bf x})$, the
microscopic rotational invariance in the order parameter space is
restored and we get back the usual $O(N)$ model with quenched
disorder. On the other hand, if all $r_i=0$ and $v\neq 0$, it
reduces to the well-known cubic anisotropic model
\cite{lubensky-book}.  Furthermore, with all $r_i ({\bf x}) =
\tilde r ({\bf x})$ and both $u,v\neq 0$, it is identical to the
model of Ref~\cite{old}. We have kept the gradient term of the
free energy functional \ref{free} spatially isotropic for simplicity. 
This, though admittedly an
idealisation, simplifies the ensuing algebra considerably {\em
without destroying the effects of chirality of the
disorder}~\cite{jacques-book}. The cubic anisotropic terms
reflect any possible breakdown of rotational invariance of the order parameter.
The relative relevance (in a RG sense) between $u$ and $v$ has been addressed by
perturbative RG calculations, see, e.g., \cite{lubensky-book}: For
$N<4$, the scaling properties are described by the stable isotropic
FP $v_R=0,\,u_R>0$, with suffix $R$ referring to renormalized
quantities, and the associated scaling exponents at the critical
points are identical to that for the usual $O(N)$ model (with bare
$v=0$)~\cite{lubensky-book}. In contrast, for $N>4$, the cubic
anisotropy is a {\em relevant perturbation} on the isotropic FP and
the system cross over to the {\em cubic anisotropic} FP with
$u_R>0,\,v_R>0$. In addition, the system can display {\em
fluctuation induced first-order transition} due to the
Coleman-Weinberg mechanism for a range of (bare) values of the
coupling constants~\cite{amit}. With this background and in order to
include the effects of spatial reflection symmetry breaking in the
disorder distribution, we assume the fluctuations in $T_c^L$,
$r_i({\bf x})$ in our model to be Gaussian distributed with
variances
 \bea \langle r_i({\bf x})r_j(0)\rangle_{av} =
2D\delta_{ij}\delta({\bf x}) + a_{ij} [\hat{D}\delta({\bf x}) +
\tilde{D}({\bf x})], \label{disvar}\eea
 with the matrix $a_{ij}$ having a structure of the form $a_{11}=0=a_{22}$ 
and $a_{12}=1=a_{21}$; function $\tilde D({\bf x})=-\tilde D({\bf -x})$,
 $D$ is a positive constant and $\hat D$ is a constant that can be both positive and
 negative. The symbol $\langle ...\rangle_{av}$ represents averages 
over the chosen Gaussian disorder distribution. A non-zero
 $\tilde D({\bf x})$ thus introduces breakdown of parity in the
 system.  Since
$\tilde D({\bf x})$ is an odd function of $\bf x$, its Fourier
 transform must be imaginary and odd in Fourier wavevector $\bf q$, such that the Fourier
 transform of $\tilde D({\bf x})$ is given by $i D_\times({\bf q})$, where $ D_\times({\bf q})$
 is a real odd function of $\bf q$: $ D_\times({\bf q})=- D_\times({\bf -q})$. In order for
 $\tilde D({\bf x})$ to be {\em equally relevant} (in a RG sense) with $D\delta ({\bf x})$ and $\hat D\delta ({\bf x})$,
 i.e., for them to have the same physical dimensions, $D_\times({\bf q})$ should
 have the same $q$-dependence (in a power counting sense) as $D$ or $\hat D$. Since, $D$ and $\hat D$ are just constants, independent of $q$,
 $ D_\times({\bf q})$ must not depend on the magnitude of $\bf q$. Therefore, we further define an amplitude $D_\times^2 =  D_\times({\bf
 q})  D_\times({\bf q})$ and a dimensionless ratio $N_\times =
 (D_\times/\hat D)^2\geq 0$. Clearly, $\tilde D({\bf x})$ is an odd function of $\bf x$ that has the same
 dimension $1/L^d$ ($L$ is a length) as the $\delta$-function $\delta^d({\bf x})$ in $d$-dimension. It is easy to work out an
 explicit representation of $\tilde D({\bf x})$ in $1d$: Writing $\tilde D(x)$ as the inverse Fourier Transform of
 $D_\times (q)$ where $q$ is a one-dimensional Fourier wavevector, we find
 \begin{eqnarray}
 \tilde D(x) = i\int_0^\infty dk D_\times\; [\exp (ikx)-
 \exp(-ikx)]\sim D_\times/x.\label{crossform}
 \end{eqnarray}
 {Although $1/x$ has a range longer than $\delta (x)$, it has the same physical
 dimension and hence scales the same way as $\delta(x)$ under rescaling of 
space, i.e., $x$. This paves the way for different
 elements in (\ref{disvar}) to compete in an RG sense. The $d$-dimensional 
analogue of the $1d$ form of $\tilde D (x)$ above is rather complicated.
 Nevertheless, it should generally be of the form $1/r^d$ on general dimensional 
ground in one hemisphere, with a change in sign in the other hemisphere. We then ask:
 What are the scaling properties of our model?}
 Our heuristic
 arguments and detailed calculations below reveal that $N_\times$ may be
 varied
 to tune the emerging critical scaling behaviour continuously.
 Before we discuss our results in details below, a few words about the interpretation of the structure of our
 model in the context of possible physically realizable examples are
 in order: {The mixing of order parameter indices and
spatial dependence in (\ref{disvar}), although not allowed in the
usual spin models, is consistent with the Frank free energy for
nematic liquid crystals~\cite{jacques-book} where the director field
is defined in the coordinate space. A chiral disordered material has
both positional and orientational disorder. For instance, in a
chiral aerogel the positional disorder is related to the local
density fluctuations of the aerogel pores, where as the
orientational disorder reflects the randomness in the orientation of
the pores.  The latter one, represented formally by a quenched
vector field~\cite{ben}, may in general have a parity breaking
variance, and should couple with the fields, {e.g., the nematic
director field in the N-A or the displacement fields in the A-C
transitions. While detailed form of such couplings are
model-specific and are not necessarily as simple as we have in
(\ref{disvar}), these}, in-principle, should generate disorder
distributions that breaks symmetry under inversion. A simple choice
as (\ref{disvar}), despite its limitations, suffices for our
purposes here. In general, the distributions of the two types of
disorders may not have any simple relation as they may occur
independently.}

\section{ Critical scaling exponents}
\label{res}

Our model (\ref{disvar}) without any disorder (all $r_i=0$
identically) displays standard order-disorder transition through a
second order critical point (in addition to a fluctuation induced
first order transition) as discussed above. To get an idea about the
possible macroscopic effects of disorder on the pure system
properties, it is instructive to first consider the prediction of
the Harris criteria \cite{harris} for the present model.
 In order to retain
the effects of the parity breaking part of the disorder variance
while constructing  the Harris criterion, we formulate it as given 
below. While here we closely follow the derivation of the
standard Harris criterion and the notations as in Ref.~\cite{wh}, we
nevertheless rephrase the details here again for the sake of
completeness. To this end, we divide the system into subsystems of
linear dimension $\xi$, where $\xi$ is the correlation length at
that temperature of the corresponding pure system. The idea is to
find out if the variation of the critical temperature of these
regions of size $\sim\xi$ becomes negligible as $T\rightarrow T_c$.
Since the spins are expected to be correlated and on average aligned
for up to a distance $\sim\xi$, the transition temperature for the
$i$th component $\phi_i$ of the order parameter field of a region of
size $\xi$ may be defined as the average of $T_{ci}^L({\bf x})$ over
that region. We define reduced temperature $t=(T-T_c)/T_c$ and local
reduced temperature $t_i ({\bf x})= \overline r_i ({\bf
x})/T_c,\,\overline r_i ({\bf x})= r_0 + r_i ({\bf x})$ for the
$i$-th component of the order parameter field, we have $\langle
t_i({\bf x})\rangle_{av} = t$. Further, as defined in
Ref.~\cite{wh}
\begin{equation}
t_{iV}=\frac{1}{V}\int d^dx t_i ({\bf x}) \label{tiv}
\end{equation}
is the effective reduced temperature of a region $V=\xi^d$. Note
that we have formally allowed an effective reduced temperature
$t_{iV}$ for the $i$-th component $\phi_i$ of the order parameter
field. This is consistent with the fact that our model allows for
order parameter component dependent effective reduced temperature 
in the free energy functional \ref{free}. 
The variance $\Delta_{ij}$ of $t_{iV}$ is defined as
\begin{eqnarray}
\Delta_{ij}&=&{\langle t_{iV} t_{jV}\rangle^c_{av}}\nonumber
\\ &=&\frac{1}{V^2}\int_V d^dx \int_V d^dy {\langle
t_{i}({\bf x})t_{j}({\bf y}) \rangle^c_{av}}\nonumber \\
&=&\frac{1}{T_c^2}\frac{1}{V^2}\int_V d^dx\int_V d^dy g_{ij}({\bf
x-y}),
\end{eqnarray}
where $g_{ij}({\bf x-y}) = \langle r_i ({\bf x}) r_j({\bf
y})\rangle_{av}$, as given in Eq.~(\ref{disvar}). Here, a
superscript $c$ refers to the connected part of the variance. Thus
\begin{eqnarray}
\Delta_{ij}&=&\frac{1}{T_c^2}\frac{1}{V^2}\int_V d^dx\int_V d^dy
\left(2D\delta_{ij}\delta({\bf x-y}) + a_{ij} [\hat{D}\delta({\bf
x-y}) + \tilde{D}({\bf x-y})]\right).\label{deltadef}
\end{eqnarray}
It is clear from (\ref{deltadef}) that the contribution from
the odd parity part of the disorder variance to the variance
$\Delta_{ij}$ vanishes owing to the odd parity of $\tilde D ({\bf
x})$. In order to capture the effect of the parity breaking part of
the disorder variance (i.e., non-zero $D_\times$), we modify
Eq.~(\ref{deltadef}) to
\begin{eqnarray}
\Delta_{ij}&=&\frac{1}{T_c^2}\frac{1}{V^2}\int_V d^dx\int_V d^dy
\left(2D\delta_{ij}\delta({\bf x-y}) + a_{ij} [\hat{D}\delta({\bf
x-y}) + |\tilde{D}({\bf x-y})|]\right).\label{deltadef1}
\end{eqnarray}
Alternatively, one may restrict the domain of integrations
above to hemisphere having a single signature of the parity breaking
part in the disorder variance. While the above modification, in
terms of considering the absolute value of $\tilde D({\bf x-y})$, is
admittedly {\em apriori} designed to capture non-zero contributions
from the parity breaking part of the disorder distribution, this
does not alter the power counting in the integral in
(\ref{deltadef}). For $D_\times=0$ this modification still leads to
 the well-known Harris criterion~\cite{harris,wh}. With our
modification, therefore, $\Delta_{ij}$ is expected to have a
contribution proportional to $|D_\times|$ or $D_\times^2$. From
(\ref{deltadef1}) we note that,
$\Delta_{ij}$ will have a part $\sim \xi^{-d}$ coming from $D$ and
$\hat D$ in (\ref{disvar}). In addition, there should a part $\sim
\xi^{-d}\ln\xi$ (this may be shown explicitly in $1d$ with $\tilde
D(x)\sim 1/x$) coming from the parity breaking part in
(\ref{disvar}). Proceeding as in Refs.~\cite{harris,wh}, we then
conclude that disorder is relevant as long as the specific heat
exponent of the corresponding pure system $\alpha>0$. This condition
is same as the usual Harris criteria~\cite{harris,wh}. Consider now the fact that the borderline of relative relevance (in a RG sense)
between the short-ranged and long-ranged disorder (with a variance $|x|^{−a}$) is determined
by the condition $a = d$ (\ref{crossform}), which yields a logarithmic contribution to the analogue of
$\Delta_{ij}$ in Ref. \cite{wh}. Thus looking at the logarithmic dependence associated with the variance
$\Delta_{ij}$ above, it appears that the present model is at the borderline between ($\delta$-correlated)
short-ranged and long-ranged disorder. Furthermore, for relevant long-ranged disorder the
scaling exponents depend explicitly on $a$. Since we can write $|x|^{−a} \sim |x|^{−d+\delta} 
\sim \delta |x|^{−d} \ln x$ with $a = d - \delta$, $\delta \rightarrow 0$, 
$\delta$ should appear as a control parameter in the scaling exponents.
Hence, drawing on the analogy between $\Delta_{ij}$ as above and the corresponding expression in
Ref. \cite{wh}, and comparing with (\ref{disvar}), amplitude 
$D_\times$ (equivalently $N_\times$) should appear as a tuning
parameter. 
Thus, any correction to the
critical exponents due to $D_\times$ must be at least
$O(D_\times)^2$ [and hence $O(N_\times)$], since our perturbative
calculations given below should be analytic in $D_\times$ (or $N_\times$), 
where $N_\times$ appears as an expansion parameter. While
our arguments above are of heuristic nature and do not constitute a
rigorous proof, they are indicative of non-trivial behaviour with
finite $D_\times$; our detailed RG calculations below confirm this
qualitative physical picture.

In order to systematically investigate the properties of systems
with quenched disorder it is required to average the free energy
over the disorder distribution. This can be conveniently done using
the well-known replica method \cite{replica} 
We start with the partition function for
the free energy functional (\ref{free})
 \bea Z=\prod_{i=1}^N
Tr_{\phi_i}\exp[-\beta\mathcal{F}\{\phi_i\}]. \eea Then the free
energy averaged over the disorder distribution can be written as
\bea && F \equiv -\langle\ln Z\rangle_{av} = \lim_{m\rightarrow
0}\left[\frac{\langle Z^m\rangle_{av}-1}{m}\right] \nonumber \\
&&= \lim_{m\rightarrow 0} \left\langle\left[
\frac{\prod_{\alpha=1}^m\prod_{i=1}^N Tr_{\{\phi_i^\alpha\}}
\exp[-\beta\mathcal{F}\{\phi_i^\alpha\}]-1}{m}\right]\right\rangle_{av}.
\eea
 Here,  $\alpha=1,2,....,m$ are the replica indices and
$\{\phi_i^\alpha\}$ represents $m$ replications of the order
parameters $\phi_i$.
 The corresponding $m$-replicated disorder averaged partition
 function is given by (we set $k_BT=1$, where $k_B$ is the Boltzmann
 constant)
  \bea &&\langle Z^m\rangle_{av} =
\prod_{\alpha=1}^m\prod_{i=1}^N \int
\mathcal{D}\phi_{i\alpha}\nonumber \\ &&\exp\left[\int d^dx
\left\{{r_0 \over 2}\sum_{i=1}^N\sum_{\alpha=1}^m\phi_{i,\alpha}^2 +
{1 \over 2}\sum_{i,\alpha}(\nabla\phi_{i\alpha})^2
\right\}\right]\nonumber \\ && \times\exp\left[\int d^dx\left\{
u\sum_{\alpha=1}^m(\sum_{i=1}^N\phi_{i\alpha}^2)^2+ v \sum_{i=1}^N
\sum_{\alpha=1}^m\phi_{i\alpha}^4\right\}
\right]\nonumber \\
&&\times\exp\left[-D\int
\sum_{i=1}^N\mathop{\sum_{\alpha,\beta=1}^m}_{\alpha\neq\beta}
\phi_{i\alpha}^2\phi_{i\beta}^2 -\hat D\int
\mathop{\sum_{i,j=1}^N}_{i<j}\mathop{\sum_{\alpha,\beta=1}^m}_{\alpha\neq\beta}
\phi_{i\alpha}^2\phi_{j\beta}^2\right]\nonumber \\ &&
\times\exp\left[ -\int \mathop{\sum_{i,j=1}^N}_{i<
j}\mathop{\sum_{\alpha,\beta=1}^m}_{\alpha\neq\beta}\phi_{i\alpha}^2(x)
\tilde{D}(x-x') \phi_{j\beta}^2(x')\right]. \label{partfunc} \eea
 Here
$\alpha,\beta=1,2,....,n$ are the replica indices. Nonlinear terms
with coupling constants $D,\,\hat D$ and $\tilde D$ involve fields
with different replica indices; thus, these terms lead to mixing of
replica indices. These terms in Eq.~(\ref{partfunc}) originate due
to the averaging over the disorder distribution. Our purpose is to
calculate the scaling exponents $\eta$ (anomalous dimension) and
$\nu$ (correlation length exponent) near the second order phase
transition, which are formally defined through the relation $\langle
\phi_i({\bf r})\phi_j(0)\rangle \sim r^{2-d-\eta}
f^\phi_{ij}(r/\xi)$ where $\xi\sim |T-T_c|^{-\nu}$ is the
correlation length; $f_{ij}^\phi$ is a dimensionless scaling
function and averages $\langle..\rangle$ are thermal averages
to be obtained from the disorder averaged free energy functional.
Anomalous dimension $\eta$ describes the spatial scaling of the
correlation function at $T_c$, where as $\nu$ describes the
divergence of $\xi$ as $T$ approaches renormalized $T_c$. If we take
only the terms quadratic in $\phi_i^\alpha$ in Eq.~(\ref{partfunc}),
the model can be solved exactly. When the nonlinear terms are
present, exact solutions are ruled out. The RG framework provides a
systematic method to extract scaling behaviour in a nonlinear
theory. The detailed methods of the RG framework is well-documented
in literature, see, e.g., Ref.~\cite{amit,zinn}; see also
Ref.~\cite{lub} for applications of RG in a disordered system. The
presence of the nonlinear terms requires expanding the different
vertex or correlation functions in powers of the coupling constants.
Near the critical point the perturbative corrections diverge leading
to failure of na\"ive perturbation expansion. The perturbative
corrections are represented by using the standard Feynmann diagrams
\cite{zinn}. We use a minimal subtraction scheme together with 
the dimensional regularisation scheme to evaluate the diagrams, where only the
diverging parts of the diagrammatic corrections are obtained in
inverse powers of $\epsilon=d_c-d$. Here $d_c$ is the upper critical
dimension, at which the relevant coupling constants become
dimensionless. For the present model, $d_c=4$ for all the
nonlinearities~\cite{upp-crit}. Thus all of them are {\em equally}
relevant. We define the renormalized coupling constants via the
renormalization $Z$-factors
$u_R=Z_uu\mu^{-\epsilon}A_d,\;v_R=Z_vv\mu^{-\epsilon}A_d,\;
D_R=Z_DD\mu^{-\epsilon}A_d,\; {\hat D}_R=Z_{\hat D}\hat D
\mu^{-\epsilon}A_d,$
 where $\mu$ is an arbitrary
 momentum scale  and $A_d={1 \over 2^d\pi^{d/2}}$. The
$Z$-factors, defined to absorb the divergences, are used to obtain
the RG equation for the correlation or vertex functions, which yield
the scaling exponents at the different RG FPs.  The FPs are formally
given by the zeros of the RG $\beta$-functions
$\beta_a=\mu\frac{\partial}{\partial \mu} a,\,a=u,v,D,\hat D$ in the
present model. The $\beta$-functions are found to be (there are no
corrections to $D_\times$ up to two loop orders)
 \bea \beta_u &=& [ -\epsilon + 48u_R + 8u_R (N+2) +48v_R-
32D_R \nonumber \\ &-& 32\hat D_R]u_R, \nonumber \\
\beta_v &=& [-\epsilon + 96u_R + 72v_R- \frac{64u_RD_R}{v_R}
\nonumber \\ &+&\frac{32u_R\hat D_R}{v_R} - 96D_R]v_R,
\nonumber \\
\beta_D &=& [-\epsilon +48(u_R+v_R) -64D_R \nonumber \\ &+&
8(N-1){u_R{\hat D}_R \over D_R}]D_R, \nonumber \\
\beta_{\hat
D} &=& [ -\epsilon + 48(u_R+v_R) -32D_R + 32{u_RD_R \over {\hat
D}_R}\nonumber \\ &+& 16(N-2)u_R -16{\hat D}_R - 16{D_\times^2 \over
{\hat D}_R}]{\hat D}_R. \label{betas}\eea
 The RG FPs are given by the zeros of the $\beta$-functions (\ref{betas}). It is useful to first consider the case
with $D_\times=0$. The FPs are
\begin{itemize}
\item Gaussian FP: $u_R=0,v_R=0,D_R=0,\hat D_R=0$,
\item Heisenberg FP: $u_R={\epsilon \over 8(N+8)},v_R=0, D_R=0,\hat
D_R=0$,
\item Cubic anisotropic (CA) FP: $u_R={\epsilon \over
24N},v_R={(N-4)\epsilon \over 72N}, D_R=0,\hat D_R=0$, \item Random
isotropic (RI) FP: $u_R=\frac{\epsilon}{32(N-1)},
v_R=0,D_R=\frac{(4-N)\epsilon}{128(N-1)},\hat
D_R=\frac{(4-N)\epsilon}{64(N-1)}$,
\item Random cubic (RC) FP:
$u_R={\epsilon \over 48(N-2)},\,v_R=\frac{\epsilon}{144}
\frac{N-4}{N-2},\,D_R=\frac{\epsilon}{192}\frac{4-N}{N-2},\,\hat D_R
= 2D_R = \frac{\epsilon}{96}\frac{4-N}{N-2}$.
\end{itemize}
Among the coupling constants, only $D_R$ must be non-negative in
order to be physically meaningful. Thus, both RI and RC FPs may
exist only for $4\geq N$~\cite{rc1}. For $N>4$ the disorder will not
be relevant any more and the system will be described by the pure
system FPs. Although even with $D_\times =0$ microscopically
our model here is different from (and a slight generalisation of)
that in Ref.~\cite{old} in having two parameters $D$ and $\hat D$
denoting variances of the disorder distributions, note that at the
RI FP, $\hat D_R=2 D_R$. This holds at the RC FP as well, although
the overall system is no longer isotropic (due to a non-zero $v_R$) 
in the order parameter space.
Thus, generically, $\hat D_R = 2 D_R$ at the FPs that depend
upon the disorder. This shows the redundance of having two
independent parameters $D$ and $\hat D$ in a RG sense, and
demonstrates the robustness of the results of Ref.~\cite{old}. We,
however, shall see below that independent $D$ and $\hat D_R$ are
required when $D_\times\neq 0$. Unsurprisingly, our results on
the FP values of the coupling constants at the RC FP above (with
$\tilde D ({\bf x})=0$) match with those of Ref.~\cite{old}.
In fact, with $\hat D_R=2D_R$ and $\tilde D({\bf x})=0$, the
$\beta$-functions (\ref{betas}) exactly correspond to the recursion
relations for the coupling constants in Ref.~\cite{old}. At the
Gaussian, Heisenberg and CA FPs, the exponents are
well-known~\cite{lub}. With the RG FPs available, the different
scaling exponents may now be obtained by using standard procedures:
One begins by calculating the
 the RG $Z$-factors $Z_{r}$ and $Z_\phi$, defined via renormalised
$r_R=Z_r r_0$ and $\phi_R=Z_\phi \phi$. The $Z$-factors $Z_r$ and
$Z_\phi$ are obtained from the one- and two-loop Feynmann diagrams
by using a minimal subtraction scheme together with dimensional
regularisation. The corresponding Wilson flow functions
$\gamma_r=\ln Z_r$ and $\gamma_\phi = \ln Z_\phi$ then immediately
yield the exponents $\nu$ and $\eta$ respectively. For further
details we refer the reader to Refs.~\cite{amit,zinn}. The scaling
exponents at the RI FP are modified by the quenched disorder and are
given by
\begin{eqnarray}
\eta &=&  {(N+2)\epsilon^2 \over 32(N-1)^2} - {(4-N)(N+2)\epsilon^2
\over 64(N-1)^2} + {(N-4)^2\epsilon^2 \over 256(N-1)^2}, \nonumber \\
{1 \over \nu} &=& 2-{3N\epsilon \over 8(N-1)}. \label{expo1}
\end{eqnarray}
These are identical to those at the random FP in Ref.~\cite{lub}.
The remaining critical exponents may be obtained from the exponent
expressions (\ref{expo1}).
We perform linear stability analyses around the RI and RC FPs; we
consider only $N<4$, for which these FPs are physically
meaningful~\cite{gauss}. Since $2D_R=\hat D_R$ is maintained at both
RC and RI FPs, we are concerned with linear stability in the
$(u_R,\,v_R,\,D_R)$ space. Further, we concern ourselves with second
order phase transitions only. First, the RI FP: We find for the
eigenvalues of the linear stability matrix
\begin{equation}
\Lambda=\epsilon,\,\frac{(4-N)\epsilon}{4(N-1)},\,\frac{(4-N)\epsilon}{4(N-1)}.
\end{equation}
Thus, for $N<4$ all the eigenvalues are positive, and hence stable
in all the three directions. The discussions of the
stability of the RC FP in Ref.~\cite{old} directly apply here as
well. For the sake of completeness, we show the eigenvalues up to
$O(\epsilon)$ at the RC FP
\begin{eqnarray}
\Lambda&=& \epsilon, \, {(N-4)\epsilon \over 6(N-2)},\,{(4-N)\epsilon
\over 6(N-2)}.
 \end{eqnarray}
Thus, two of them diverge at $N=2$. It is thus generally expected
that flow lines from near the RC FP flow to the (stable) RI FP.
See Ref.~\cite{old} for detailed discussions on this.


Having known the FP values of the coupling constants for $D_\times
=0$ or $N_\times =0$, we now obtain the corresponding values when
$N_\times >0$. For simplicity of the ensuing algebraic
manipulations, we find corrections to the ($N_\times =0$) FP values
of the coupling constants up to $O(N_\times)$, assuming a small
$N_\times$. Thus the finite $N_\times$ FP values, as we write down
below, are not going to be quantitatively accurate for $N_\times
\sim O(1)$. For $D_\times=0$, the zeros of the
$\beta$-functions $\beta_D$ and $\beta_{\hat D}$, as given in
Eqs.~(\ref{betas}), yield $\hat D_R= 2D_R$ generically at the RG
FPs. This then, in turn, leads to $\hat D_R=0= D_R$ as {\em a FP
solution}, regardless of its linear stability properties. Thus, pure
system properties may be restored in the large scale provided $\hat
D_R=0= D_R$ is a stable FP. However, when $D_\times \neq 0$, two
observations may be made immediately: (i) $\hat D_R \neq 2 D_R$ and
(ii) $\hat D_R=0$ and $D_R=0$ are no longer solutions of
$\beta_{\hat D}=0$ and $\beta_{D}=0$. Thus, pure system behaviour
is not expected to be observed. We obtain modifications to both RI
and RC FPs up to $O(N_\times)$ separately. First the modified RI FP:
After some straight forward algebra, we find
\begin{eqnarray}
u_R^*&=&\frac{\epsilon}{32(N-1)}\left[1+\frac{(4-N)
N_\times}{2(4+N)}\right],\label{ufix}\\
v_R^* &=&\frac{\epsilon(4-N)N_\times}{16(4+N)(N-1)},\label{vfix} \\
D_R&=&\frac{\epsilon(4-N)}{128(N-1)}\left[1+\frac{(16-N)}{2(4+N)}N_\times\right],\label{dfix} \\
\hat D_R &=&\frac{\epsilon (4-N)}{64 (N-1)}
\left[1+\frac{(N+8)N_\times}{2(4+N)}\right].\label{dhatfix}
\end{eqnarray}
Here, $N_\times= {D_\times^2 \over {\hat D}_R^2}\geq 0$ is a
dimensionless parameter. Notice that: (i) $v_R$ picks up a small
non-zero value $O(N_\times)$ and (ii) $\hat D_R\neq 2D_R$. Thus the
modified RI is {\em no longer isotropic}; however, the departure
from isotropicity is {\em small}, $O(N_\times)$. It is clear from
Eqs.~(\ref{ufix}-\ref{dhatfix}) that for $1<N<4$ the FP values are
physically meaningful, e.g., $D_R$ is positive definite, for any
values of (small) $N_\times$; we, therefore, consider only the range
$1<N<4$ below. Corrections to $O(N_ \times)$ for the RC FP may also
be obtained, which we do not show here explicitly; see
Ref.~\cite{old} for the details in this context. For small
$N_\times$ and $2<N<4$ this modified RC FP remains unstable. The
general picture of the second order phase transition remains
unchanged at such finite but small $N_\times$. When $N_\times \sim
O(1)$, the FPs are expected to be substantially modified, which we
do not discuss here. We are unable to comment about the physical
picture for $N>4$ on the basis of the FPs given by
(\ref{ufix}-\ref{dhatfix}), since $D_R<0$ for $N>4$ from
Eq.~(\ref{dfix}). 

The critical exponents $\eta$ and $\nu$, which may be evaluated
following the procedure outlined above, at the modified RI FP are:
\bea \eta &=& {(N+2)\epsilon^2 \over 32(N-1)^2} -
{(4-N)(N+2)\epsilon^2 \over 64(N-1)^2} + {(N-4)^2\epsilon^2 \over
256(N-1)^2} \nonumber
\\&-& {(4-N)^2\epsilon^2 \over 256(N-1)^2}N_\times
, \label{eta} \\
{1 \over \nu} &=& 2-{3N\epsilon \over 8(N-1)} - {3(4-N)\epsilon
\over 16(N-1)}N_\times, \label{nu} \eea
 As usual, the critical exponent expressions (\ref{eta}) and (\ref{nu}) yield all
 other critical exponents. The scaling exponents given in
 (\ref{eta}) and (\ref{nu}) constitute the principal results of this
 work.
 From Eq.~(\ref{eta}),  for $N_\times=0$, $\eta$ is always
positive (we consider $1<N<4$). For small $N_\times\neq 0$, $\eta$
remains positive, but its value decreases. Thus the correlation
function for $\phi_i$ decays more rapidly with spatial separation
for $N_\times=0$ than for $N_\times\neq 0$. It apparently suggests
that $\eta$ can be brought to zero for sufficiently large
$N_\times$; however, since our expression (\ref{eta}) are valid for
small $N_\times$ only, we cannot say anything conclusively when
$N_\times$ is large$\,\sim O(1)$. Similarly, from Eq.~(\ref{nu}),
$\nu^{-1}$ remains positive for small value of $N_\times$ and it has
a larger value when $N_\times > 0$ than for $N_\times= 0$. Since
$\nu$ is an explicit function of $N_\times$, the measured diverging
correlation length $\xi\sim |T-T_c|^{-\nu}$ will depend on the value
of $N_\times$; $\xi$ for $N_\times > 0$ is larger than that for
$N_\times =0$. Furthermore, notice that $\nu$ picks up an
$N_\times$-dependent part at $O(\epsilon)$, where as the same for
$\eta$ appears at $O(\epsilon^2)$. Thus, experimental detection of
any $N_\times$ dependence will be revealed much more clearly in
measurements of the correlation length near renormalized $T_c$.


What could be an upper bound of $N_\times$? By demanding positivity
of the eigenvalues $\Lambda_e$ of the disorder variance matrix, we
can enforce a bound on $ D_\times$ (or, equivalently, an upper limit
on $N_\times$). Eigenvalues $\Lambda_e$ in general depend both on
$N$, the number of order parameter components, and $D,\,\hat D,\,
D_\times$. We take $N=2$ as a specific example. The variance matrix
$M$ of the disorder distribution is given by
\begin{eqnarray}
M=\left(\begin{array}{cc}
2D & \hat{D} + i D_\times\\
\hat{D}-i D_\times & 2D
\end{array}
\right).\end{eqnarray}
 The corresponding eigenvalues are $\Lambda_e = 2D\pm \sqrt {(\hat D^2
 +  D_\times^2)}\geq 0$. This gives a bound on the off-diagonal
 elements of the variance matrix $M$. This is consistent with the limit $D_\times=0$, where $\hat D_R = 2D_R$ at the FP. Similar exercises may be under
 taken for higher values of $N$, which we do not discuss here.
 Finally, our claim of continuously varying universal properties
 rests on the possible continuous variation of $ D_\times$, and hence of
 $N_\times$. We have shown, in our low order (two-loop) renormalized
 perturbation theory, that there are no fluctuation corrections to
 $D_\times$. We believe this holds to any order in the perturbative
 expansion. To see this notice that any non-zero fluctuation
corrections to $ D_\times({\bf q})$ must be an odd function of the
its wavevector argument. In order to have such a non-zero correction
one must have an odd number of $ D_\times({\bf q})$ vertex in the
diagram. Since all internal wavevectors are integrated over, such a
contribution will vanish in the limit of vanishing external
wavevector. Thus any putative diagrammatic corrections to $
D_\times({\bf q})$ vanishes and hence $N_\times$ should appear as a
dimensionless marginal operator to any order in perturbation.
This is technically similar to a marginal operator that exists in
the models of Refs.~\cite{noise,noise1}. Numerical verification of
our results on equivalent lattice-gas models requires generation of
$N$ stochastic functions having variances as given by
(\ref{disvar}). This may be conveniently done by following the
method outlined in Ref.~\cite{noise1} (see also Ref.~\cite{stanley}
for a general discussion on related issues).

\section{Summary and outlook}
\label{conclu}

In summary, thus, we have proposed and
 {studied} a  {variant of the} classical cubic
anisotrpic $O(N)$ model with short ranged quenched disorder
 {having a parity breaking part, with a strength
parametrised by $ D_\times$ (equivalently by $N_\times$), in its
variance, as a reduced minimal model to study the effects of
quenched chiral disorder on the scaling properties of pure systems}.
For our work, we use a generalisation of the model used in
Ref.~\cite{old}.
The truly novel result from our reduced model is that the
explicit and continuous dependence of the scaling exponents on
$N_\times\neq 0$, with our results reducing to those of
Ref.~\cite{old} in the limit $N_\times =0$. These
$N_\times$-dependent results on the scaling exponents are
qualitatively new contributions made in the present work. We have
worked out the dependences of the scaling exponents up to
$O(N_\times)$ for simplicity using two-loop $\epsilon$-expansions.
{If the disorder is long-ranged~\cite{wh}, i.e., if the magnitudes
of $D,\hat D,D_\times({\bf k})$ have $\bf k$-dependent parts $\sim
k^{-y},\,y>0$, in addition to constant parts, preliminary
calculations analogous to that here reveal that a parameter analogue
to $N_\times$ will appear in the scaling exponents. Thus, the
effects of a parity breaking part appears to be quite robust.}
Assuming the disorder in our model results from a variety of
microscopic sources, the tuning parameter $N_\times$ may be
interpreted as a measure of the {\em relative concentration} of the
microscopic impurities that causes a non-zero parity breaking
variance in the resulting impurity distribution. Hence, our results
are illustrations of dependences of the scaling exponents on the
{\em relative concentration}.  {Our reduced model and the results
that follow are not directly applicable to any specific system we
discussed (e.g., the N-A transition), due to the underlying
simplicity and idealisation of the model. Nevertheless, the broad
qualitative picture that chirality of impurity distributions may be
relevant in determining the universal properties of disordered
systems is sufficiently general and expected to be observed in more
realistic and specific experimentally accessible systems. From a
general point of view, our results
open up the possibility of a new paradigm in the scaling properties of quenched chiral disorder systems.} 
Notice that our results do not
contradict the recent
 results that critical exponents in the disordered Ising model are independent of impurity
concentration along the transition line between paramagnetic and
ferromagnetic phases~\cite{ising}, since $D_\times =0$ (hence
$N_\times=0$) identically for the Ising model with $N=1$. Many
experiments tend to suggest {\em smearing of phase transition} in a
variety of disorder systems~\cite{smear}. Our results on
$N_\times$-dependent correlation length as $T\rightarrow T_c$ and,
similarly, $N_\times$-dependent {\em renormalized} $T_c$ are
reminiscent of a system with a continuous spectrum of critical
points and relevant diverging length scales (near $T_c$) and
critical points,  {and thus loosely resemble a smeared transition
(although there are no formal connections)}. Measurements on model
systems (for $N\geq 2$) with different realizations of the disorder
having different $N_\times$, corresponding to a given impurity
distribution, naturally leading to a broadening of the measured
values of the scaling exponents. {Although at present no experiments
on chiral disordered systems are available to our knowledge, we
expect such experiments (e.g., N-A or A-C transitions) should be
performed in the near future by using, say, chiral
aerogels~\cite{chiral-aero}.} {
Numerical simulations of our reduced model with appropriately chosen
disorder distributions should be useful.} From the point of view of
the notion of universality in both equilibrium and nonequilibrium
systems, our results are yet another demonstration of the important
role that breakdown of spatial parity invariance may play in
determining the universal properties. Effects of parity breakdown in
pure nonequilibrium systems have  {already} been
 {elucidated} in several examples in Refs.~\cite{noise,noise1}, where
breakdown of microscopic parity invariance is introduced by means of
a reflection invariance breaking stochastic noise correlator. In
each of those cases, universal properties depend explicitly on a
model parameter analogous to $N_\times$ here. Noting that the
disorder variance in the present study (\ref{disvar}) is symmetric
under the exchange of $(i,j)$, our model may be generalised further
by allowing for terms in the disorder variance that are
antisymmetric under exchange of $i,j$, which may separately have
even and odd parity parts. We expect the coefficient of the new odd
parity part should also appear as a tuning parameter in the critical
scaling exponents. Calculations analogous to the above may be
performed for detailed study. Lastly, it will be interesting to see
how the predictions of Ref.~\cite{ostlund} for the disordered
classical XY model in $2d$ are modified for the cubic anisotropic
$O(2)$ model at $2d$ with an impurity distribution as here. In
addition, effects of quenched chiral disorder of the type discussed
here on systems similar to Ref.\cite{old2} should be investigated.
We hope our studies here will stimulate further theoretical and
experimental studies.

\section{Acknowledgement} 
AB gratefully acknowledges partial financial support in the form of
the Max-Planck Partner Group at the Saha Institute of Nuclear
Physics, Calcutta, funded jointly by the Max-Planck-
Gesellschaft (Germany) and the Indo-German Science \& Technology Centre
(India) through the MPG-DST Partner Group programme (2009).


\begin{thebibliography}{99}
\bibitem{topo} At
$2d$, our results, specially for $N=2$, are likely to get modified
by topological defects, see, e.g., Ref.~\cite{ostlund}.
\bibitem{upp-crit} The upper critical dimension $d_c$ of a field theory is
the dimension at which the relevant coupling constant becomes
dimensionless; see, e.g., \cite{amit,zinn} for more detailed
technical discussions about $d_c$ for any model. In the present
model, all the coupling constants $u,v,D,\hat D$ of the effective
disordered averaged theory, as given by (\ref{partfunc}), are
dimensionless at dimension $d=4$. Hence, $d_c=4$ for all the
nonlinearities.
\bibitem{rc1} The RC FP diverges at $N=2$. Our calculation fails to
suggest any physical interpretation or consequence for it.
\end{thebibliography}
\end{document}